# APC waivers and Ukraine's publishing output in Gold OA journals: Evidence from five commercial publishers

**Serhii Nazarovets**[1]

[1]Borys Grinchenko Kyiv Metropolitan University, 18/2 Bulvarno-Kudriavska Str., 04053 Kyiv, Ukraine. http://orcid.org/0000-0002-5067-4498

**Abstract**
This study examines the effect of article processing charge (APC) waivers on the participation of Ukrainian researchers in fully Gold Open Access (Gold OA) journals published by the five largest academic publishers – Elsevier, SAGE, Springer Nature, Taylor & Francis, and Wiley – during the period 2019–2024. These publishers were selected because, in response to the full-scale war launched against Ukraine in 2022, all five introduced emergency 100% APC-waiver policies for Ukrainian authors. Using bibliometric data from the Web of Science Core Collection, the study analyses publication trends in Ukrainian-authored articles in fully Gold OA journals of these publishers before and after 2022. The results show a marked post-2022 increase in Ukraine's Gold OA output, particularly in journals published by Springer Nature and Elsevier. Disciplinary and publisher-specific patterns are evident, with especially strong growth in the medical and applied sciences. The findings underscore the potential of targeted support measures during times of crisis, while also illustrating the inherent limitations of APC-based publishing models in fostering equitable scholarly communication.

**Keywords:** Gold Open Access; APC waivers; scholarly publishing; Ukraine; bibliometrics

## 1. Introduction

The full-scale military invasion of Ukraine by the Russian Federation, launched on 24 February 2022, has imposed unprecedented challenges on the Ukrainian scientific community. The destruction of infrastructure, the displacement of researchers, and the escalation of economic hardship have severely disrupted academic activities and research continuity (Nazarovets & Teixeira da Silva, 2022). Among the many consequences, the restriction of financial resources for research dissemination has emerged as one of the most critical barriers, limiting Ukrainian scholars' ability to maintain their international visibility (Gaind et al., 2022).

In this context, ensuring opportunities for Ukrainian researchers to publish their work has become essential for preserving and strengthening the country's scientific capacity. However, access to publication venues, particularly in prestigious open access journals, is often contingent on the ability to pay article processing charges (APCs), which may present an additional barrier for researchers operating under severe financial constraints.

Another critical dimension influencing access to Gold Open Access (Gold OA) publishing is the growing market concentration among academic publishers. Gold OA refers to a publishing model in which scholarly articles are made immediately and freely available by the publisher upon publication, without access restrictions or subscription fees. Unlike hybrid models, Gold OA applies to entire journals, where all content is accessible to readers at no cost. As described by Laakso et al. (2011), this model has grown rapidly since the early 2000s and now encompasses a wide range of journals across disciplines and publishers, from small independent outlets to high-volume commercial platforms such as those owned by the "Big Five" academic publishers.

Recent analyses reveal that five major companies – Elsevier, SAGE, Springer Nature, Taylor & Francis, and Wiley – dominate a substantial share of the scholarly publishing market (Butler et al., 2023). These publishers are central proponents of the APC model, and their market power enables them to maintain high article processing charges across both gold and hybrid open access journals. Consequently, while APC waivers and support initiatives help to mitigate individual-level barriers, broader structural inequalities in the global academic publishing economy persist.

To better understand the nature of this challenge, it is necessary first to define the concept of APCs and examine how they create obstacles to equitable participation in Gold OA publishing.

APCs refer to the fees charged by publishers to authors (or their institutions or funders) to make their research articles openly accessible immediately upon publication. This model of financing emerged in the early 2000s, aligned with the rise of the Open Access (OA) movement and the growing demand to remove subscription barriers that restricted access to scientific knowledge (Beasley, 2016; Solomon & Björk, 2016). Initially, pioneering OA publishers such as BioMed Central and the Public Library of Science (PLOS) introduced APCs as an alternative to subscription revenue (Beasley, 2016). The underlying idea was that in a digital environment where the marginal cost of distributing content approaches zero, shifting costs from readers to authors would democratize access to research (Green, 2019).

However, although APCs successfully removed paywalls for readers, they simultaneously introduced a new barrier – the ability to publish became contingent upon the ability to pay. As Tenopir et al. (2017) observed, while many authors recognize the benefits of OA for readership and visibility, there is widespread concern that APCs present significant obstacles, especially for researchers without institutional or grant support. Klebel and Ross-Hellauer (2023) further demonstrate that APCs have stratified access to publishing opportunities, favouring researchers affiliated with well-resourced institutions and those based in wealthier countries. The introduction of APCs has thus shifted the axis of inequality in scholarly communication – from consumption (access to reading) to production (access to publishing) (Halevi & Walsh; 2021).

While average APCs for fully OA journals typically range between $1,000–$2,000 USD (Solomon & Björk, 2016; Björk & Solomon, 2014), hybrid journals often charge significantly higher fees, exceeding $3,000 USD per article (Borrego et al., 2020; Budzinski et al., 2020). Importantly, as Borrego (2023) notes, the market dominance of major publishers has allowed APCs to remain high despite expectations that increased competition would drive prices down.

Recognizing the exclusionary potential of APCs, some publishers have introduced mechanisms such as fee waivers to assist authors from low- and middle-income countries (Lawson, 2015). Nevertheless, as discussed later, such initiatives only partially mitigate the systemic inequalities introduced by APC-based models. Thus, while APCs have opened new paths for making research outputs accessible to global audiences, they have also entrenched financial barriers at the point of authorship, raising new questions about equity in academic publishing.

In response to the financial challenges that researchers from low- and middle-income countries (LMICs) often encounter, and in an effort to promote fairer participation in global scholarly discourse, various initiatives have emerged to enhance access to scientific resources and reduce the burden of publication fees. One of the most prominent examples is Research4Life, a public-private partnership established in 2002 by organizations such as the World Health Organization (WHO), the Food and Agriculture Organization (FAO), the United Nations Environment Programme (UNEP), and the World

Intellectual Property Organization (WIPO), in collaboration with over 200 international publishers (Hill, 2021; Anyaoku & Anunobi, 2014). Research4Life targets institutions in countries classified as low- and lower-middle income based on World Bank economic indicators, the United Nations Least Developed Countries list, and the Human Development Index. Institutions from eligible countries are divided into Group A (free access) and Group B (low-cost access)[1].

In addition to facilitating reading access, Research4Life partner publishers and other major academic publishers responded to the humanitarian and research challenges arising from the full-scale war in Ukraine by introducing extraordinary temporary measures. Starting from 2022, leading commercial academic publishers – including Elsevier[2], SAGE[3], Springer Nature[4], Taylor & Francis[5], and Wiley[6] – implemented full APC waivers (100% discounts) for Ukrainian corresponding authors submitting to their Gold OA journals. These emergency policies are specifically designed to support Ukrainian researchers during wartime conditions, helping to maintain their ability to publish internationally despite disruptions to institutional operations, financial systems, and research funding streams.

This study focuses specifically on Elsevier, SAGE, Springer Nature, Taylor & Francis, and Wiley publishers because they publicly announced a full APC waiver for Ukrainian corresponding authors in response to the war. These emergency policies created a unique situation that allowed for an investigation into whether the removal of APC-related financial barriers could directly affect the publishing activity of Ukrainian researchers. Other major publishers (such as MDPI and Frontiers), as well as national open access journals, are not included in this analysis, as they did not implement equivalent APC-waiver policies during the studied period.

A comprehensive study by Kaliuzhna and Hauschke (2024) analysed the broader evolution of open access publishing in Ukraine from 2012 to 2021, including the role of national journals, participation in transformative agreements, and developments in the policy landscape. However, their analysis did not examine the specific behavioural effects of APC-waiver interventions introduced by major commercial publishers during the war. In contrast, the present study narrows its focus to a defined subset of Gold OA journals published by five global publishers that implemented such waivers for Ukrainian corresponding authors in response to the crisis.

Despite the existence of initiatives such as Research4Life and the emergency APC waivers introduced in response to the war in Ukraine, no comprehensive study to date has systematically assessed the actual impact of these policies on the publishing activity of Ukrainian researchers. While free access to reading materials and waived APCs can theoretically enhance research participation, the extent to which these measures have translated into increased publication outputs in Gold OA journals remains unclear.

Moreover, it is not known whether the effects of publisher support have been uniform across disciplines, or whether differences exist between publishers in the scope and effectiveness of their waiver programs. In addition, while neighbouring countries such as Poland, Czech Republic, and Hungary did not experience wartime disruptions of comparable magnitude, comparative analysis is necessary to contextualize the trends observed in Ukraine.

---

[1] https://www.research4life.org/access/eligibility/
[2] https://www.elsevier.com/researcher/author/open-access/choice#3-research4life
[3] https://www.sagepub.com/journals/information-for-authors/publishing-options/gold-open-access-article-processing-charge-waivers
[4] https://www.springernature.com/gp/open-science/policies/journal-policies/apc-waiver-countries
[5] https://authorservices.taylorandfrancis.com/choose-open/publishing-open-access/requesting-an-apc-waiver/eligible-countries/#discounts
[6] https://authorservices.wiley.com/author-resources/Journal-Authors/open-access/article-publication-charges/waivers-and-discounts.html

To address these gaps, the present study is guided by the following research questions:

- Has the number and share of publications by Ukrainian authors in fully Gold OA journals published by the five aforementioned publishers increased after 2022, following the implementation of 100% APC waivers?
- Are there notable differences between publishers in terms of Ukrainian authors' participation?
- Are there disciplinary patterns that influence the dynamics of Ukrainian open access publishing?
- How does the trajectory of Ukrainian Gold OA publishing compare to that of neighbouring countries such as Poland, Czech Republic, and Hungary?

By answering these questions, this study aims to provide an evidence-based assessment of the real impact of APC waiver policies during crisis conditions, contributing to broader discussions on equity, resilience, and access in global scholarly communication.

## 2. Data and methods

The empirical analysis in this study is based on publication metadata extracted from the Web of Science (WoS) Core Collection database, provided by Clarivate. WoS was selected due to several key strengths that ensure robust and reliable bibliometric data collection. First, WoS covers a comprehensive and carefully curated collection of peer-reviewed journals across all major scientific disciplines, providing consistent metadata necessary for comparative analyses. Second, WoS metadata includes reliable details regarding author affiliations, types of publication access (open vs. subscription-based), and publisher information. Third, this database allows for precise identification and filtering of fully Gold OA journals, which are central to the research objectives. Given these advantages, WoS represents an optimal source for addressing the proposed research questions. The dataset included all document types labelled as Gold OA in WoS. While APCs are generally associated with research articles and reviews, the inclusion of other types reflects the scope of the publisher's OA offerings under waiver programs.

To ensure relevance and comparability of collected data, several specific criteria were applied during data extraction:

- *Publisher selection:* The analysis is limited to journals published by the five largest commercial academic publishers – Elsevier, SAGE, Springer Nature, Taylor & Francis, and Wiley. This choice reflects not only their substantial market share and influence on the APC-based publishing landscape (Butler et al., 2023), but also the fact that these were the publishers to implement 100% APC waivers specifically for Ukrainian corresponding authors in response to the full-scale Russian invasion of Ukraine in 2022. The study focuses on this unique group of publishers to examine the potential behavioural impact of these emergency waivers on Ukrainian scholarly publishing. Other publishers that host significant volumes of Gold OA articles authored by Ukrainian researchers, such as MDPI, Frontiers, or institutional publishers from Ukraine, were intentionally excluded. The reason for this exclusion is methodological: the present study aims to isolate the effects of 100% APC waivers offered by the five selected commercial publishers. Adding other publishers, such as MDPI or Frontiers, would have mixed very different editorial and pricing models, making it difficult to isolate the impact of the waiver policy itself. It is important to note that, as shown in the recent preprint by Myroslava Hladchenko (2025) a considerable share of Ukraine's APC spending occurs in venues with less transparent and sometimes suspiciously editorial policies. Including such publishers in this analysis would have introduced strong heterogeneity in editorial quality, obscuring the specific policy effects examined here.

- *Open access status:* Only fully Gold OA journals were selected. Unlike hybrid journals, fully Gold OA journals do not rely on subscriptions, thus guaranteeing unrestricted access to all published content and ensuring that authors must typically pay an APC unless eligible for a waiver. Importantly, the 100% APC waivers for Ukrainian authors introduced by the five selected publishers applied exclusively to fully Gold OA journals and did not extend to hybrid titles.
- *Publication timeframe:* Publications within a six-year period (2019–2024) were selected. This interval allows assessment of publication dynamics both prior to and after the onset of Russia's full-scale invasion of Ukraine in 2022, providing a sufficient baseline for identifying trends and shifts related to crisis-induced publisher waiver initiatives.
- *Author affiliation:* Articles were selected based on the "CU=Ukraine" country affiliation tag in WoS, which, while practical for identifying Ukrainian-linked output, does not guarantee that the corresponding or lead author was Ukrainian – particularly in internationally co-authored works. This limitation is addressed in more detail in Section 4.4.

It is important to emphasize that this study does not aim to provide a general overview of Ukraine's open access publishing landscape, nor to assess all publishers or journal types offering Gold OA. Instead, it deliberately focuses on a narrowly defined subset of publishers – those five companies that introduced full APC waivers for Ukrainian authors during the war.

To contextualize the publication dynamics observed in Ukraine, the study includes a comparative analysis with three neighbouring countries: Poland, Czech Republic, and Hungary. These countries were selected based on their geographic proximity, historically similar scholarly communication patterns, and comparable socioeconomic contexts. Crucially, unlike Ukraine, these countries have not experienced active large-scale military conflict during the analysed period, thus serving as effective regional benchmarks for understanding the unique effects of crisis conditions on publication activity in Gold OA journals.

For each selected paper, the following metadata fields were systematically collected from the WoS: year of publication; publisher name; journal title; country affiliation tag assigned to the article; scientific discipline, classified according to the OECD fields of science classification. Using the standardized OECD classification allows consistent cross-disciplinary comparisons, helping identify potential differences in publication dynamics between various scientific fields.

To conclude, while this methodological approach offers several strengths, it also entails certain limitations, as outlined in section 4.4. Nonetheless, considering the WoS's rigorous indexing standards and the influential position of the selected publishers in international scholarly communication, the results obtained through this approach serve as a reliable basis for capturing broader trends and developments relevant to the study's objectives.

## 3. Results

### *3.1. Overall publication dynamics (2019–2024)*

The annual number of publications authored by Ukrainian researchers in fully Gold OA journals published by the five largest academic publishers (Elsevier, SAGE, Springer Nature, Taylor & Francis, and Wiley) demonstrates a clear upward trend over the 2019–2024 period (Figure 1). These results refer exclusively to fully Gold OA journals published by the five commercial publishers that implemented APC waivers for Ukrainian authors during the war. While the output remained relatively stable between 2019 and 2022 – fluctuating between 344 and 380 articles per year – a sharp increase was observed starting in 2023.

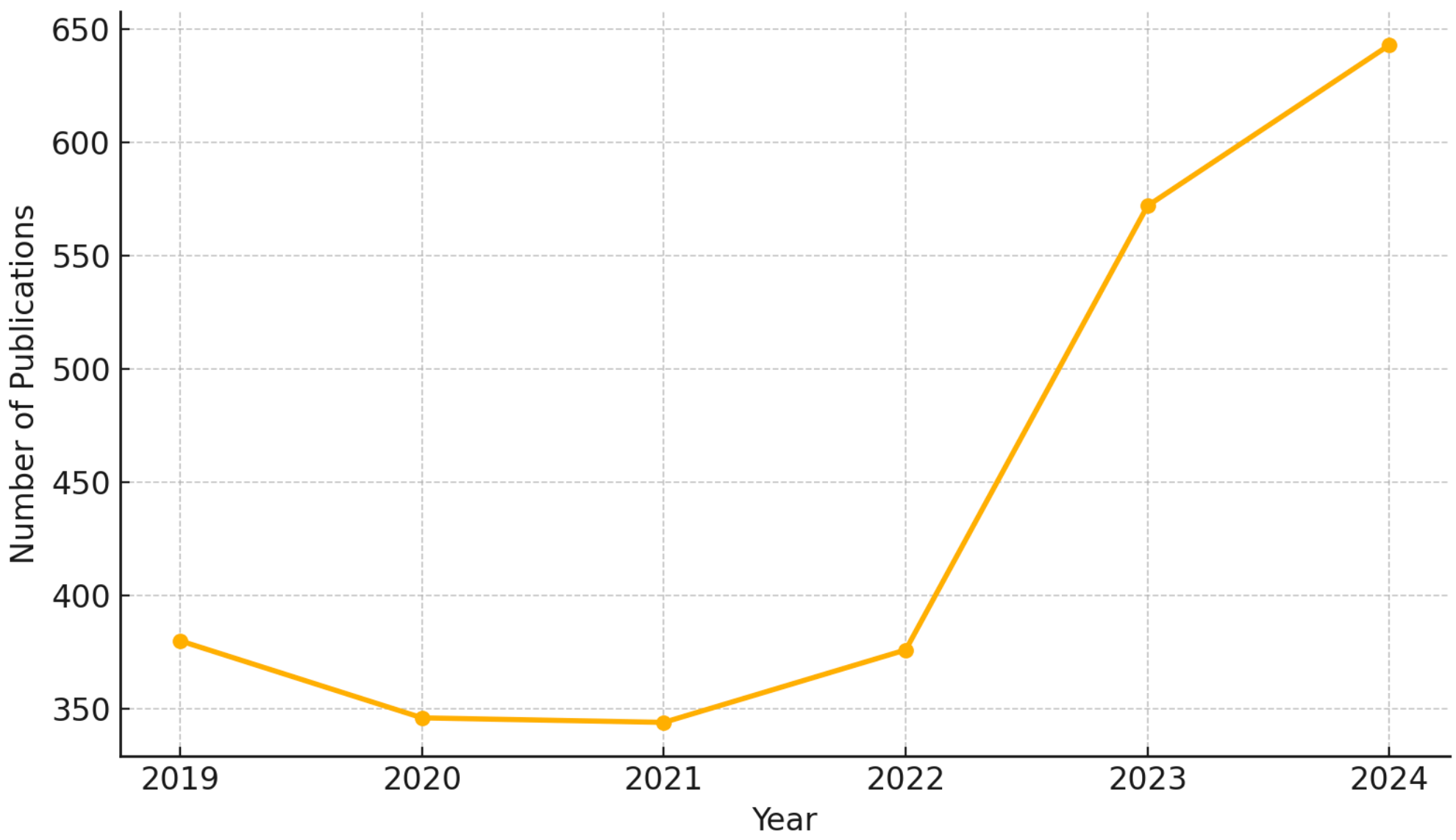

**Figure 1.** Annual number of publications by Ukrainian authors in Gold OA journals published by the five selected publishers (2019–2024)

Specifically, the number of publications rose from 376 in 2022 to 572 in 2023, representing a year-over-year increase of 52.1%. This growth continued into 2024, reaching 643 publications, which marks a further increase of 12.4% compared to 2023. These shifts contrast with the relatively stable pre-war baseline and suggest a notable change in Ukrainian researchers' publishing activity in the selected journals during the post-2022 period – potentially influenced by the introduction of full APC waivers.

*3.2. Publisher-specific trends*

A breakdown of Ukrainian-authored Gold OA publications across the five major academic publishers reveals notable disparities in their relative contributions (Table 1). Over the entire period from 2019 to 2024, Springer Nature accounted for the highest number of publications (1,204 articles), followed by Elsevier with 1,013 publications. Together, these two publishers represented the vast majority of Ukrainian Gold OA output in the selected journals.

These differences may partly reflect variations in the visibility and implementation of APC-waiver policies across publishers, as well as disciplinary alignments and author preferences. While Wiley contributed only modestly between 2019 and 2022, its role expanded significantly in 2023 and 2024, with 65 and 94 publications respectively, suggesting a growing importance in recent years. In contrast,

Taylor & Francis and SAGE maintained consistently low publication volumes throughout the period, with totals of 94 and 89 articles, respectively.

**Table 1.** Annual number of Ukrainian-authored publications in fully Gold OA journals published by the five selected publishers (2019–2024)

| **Publishers** | **2019** | **2020** | **2021** | **2022** | **2023** | **2024** | **Total** |
|---|---|---|---|---|---|---|---|
| Springer Nature | 151 | 175 | 179 | 173 | 247 | 279 | 1204 |
| Elsevier | 195 | 119 | 125 | 145 | 222 | 207 | 1013 |
| Wiley | 17 | 34 | 22 | 29 | 65 | 94 | 261 |
| Taylor & Francis | 6 | 8 | 12 | 12 | 21 | 35 | 94 |
| SAGE | 11 | 10 | 6 | 17 | 17 | 28 | 89 |
| **Total** | **380** | **346** | **344** | **376** | **572** | **643** | **2661** |

*3.3. Disciplinary profile: Top-10 research areas*

The disciplinary distribution of Ukrainian Gold OA papers published by the five selected commercial publishers from 2019 to 2024 is shown in Figure 2. As expected, Physics consistently dominates the output across all years, while other disciplines exhibit varying temporal patterns.

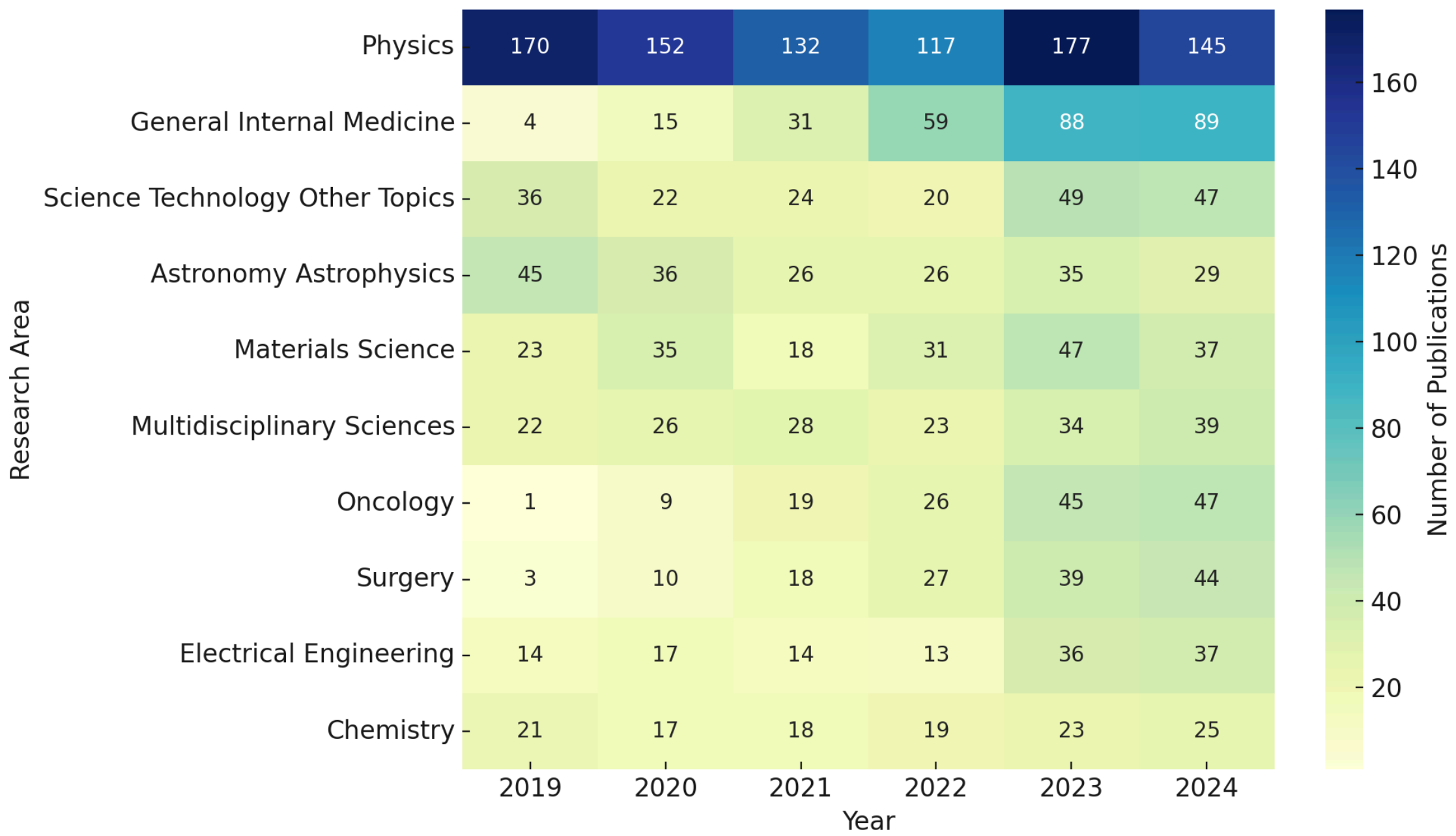

**Figure 2.** Heatmap of Ukrainian-authored publications in fully Gold OA journals by research area and year (limited to five selected publishers, 2019–2024)

While Physics has remained the dominant field throughout the period, more applied and health-related disciplines gained prominence after the introduction of APC waivers. In particular, General Internal Medicine, Oncology, and Surgery show a marked increase in publication volume during 2023 and 2024, likely reflecting shifts in scientific priorities in response to public health and wartime medical challenges. Similar trends are observed in Materials Science and Electrical Engineering, which show increasing activity in recent years – potentially linked to post-war reconstruction needs, technological resilience priorities, and improved opportunities for OA publishing made possible through APC waivers.

*3.4. Comparative perspective: Ukraine and neighbouring countries*

The indexed growth in fully Gold OA publications (limited to the five selected publishers) shows distinct national trajectories between 2019 and 2024 (Figure 3). Poland maintained the highest relative output throughout the period, followed by the Czech Republic, Hungary, and Romania. Ukraine consistently had the lowest absolute numbers, yet its relative dynamics changed markedly after 2022.

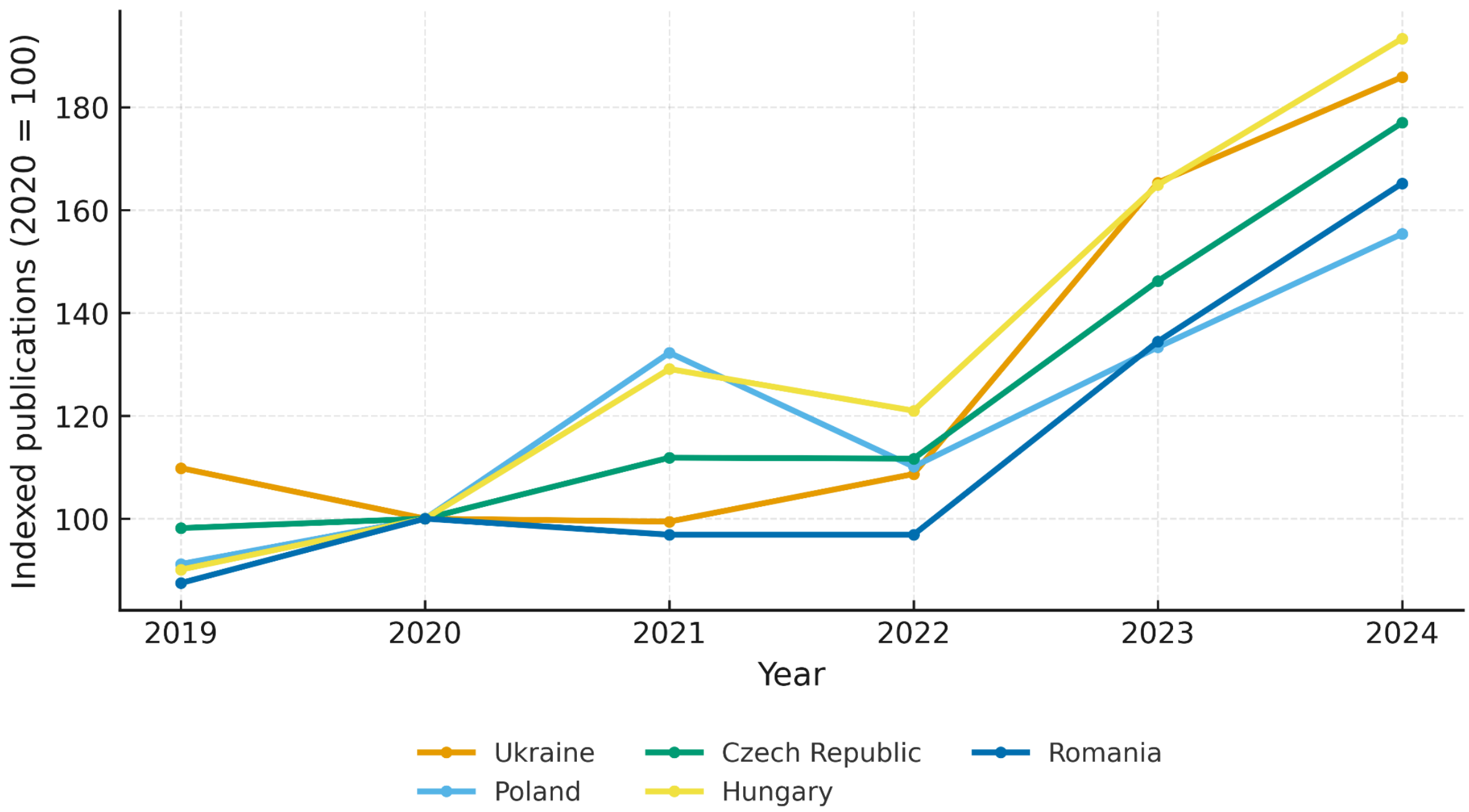

**Figure 3.** Indexed growth (2020 = 100) of publications in fully Gold OA journals (five selected publishers) for Ukraine and neighbouring countries

From 2022 onward, the Ukrainian curve rises sharply. Between 2022 and 2023, indexed output grew by more than 50 per cent and continued to increase in 2024. This upward shift, visible even after normalization, likely reflects the short-term effect of targeted support measures, most notably, the introduction of 100 per cent APC waivers by the five publishers analysed in this study. These waivers temporarily reduced financial barriers for Ukrainian authors seeking to publish in reputable Gold OA journals.

The cross-country comparison nevertheless requires caution. Countries such as Poland or the Czech Republic possess larger and better-funded research systems, which naturally generate higher publication volumes. The indexed trends therefore illustrate proportional change rather than total capacity. Further normalization by the number of researchers or R&D spending would allow a more precise assessment of relative engagement in Gold OA publishing, but this was beyond the scope of the present study.

## 4. Discussion

*4.1. Interpreting the rise in publications in Gold OA journals*

The analysis reveals a substantial increase in the number of publications by Ukrainian researchers in fully Gold OA journals published by the five academic publishers – Elsevier, SAGE, Springer Nature, Taylor & Francis, and Wiley – offered full (100%) APC waivers for Ukrainian authors after 2022. This shift departs notably from the relatively stable publication output observed between 2019 and 2021, and appears temporally aligned with the onset of Russia's full-scale invasion of Ukraine in early 2022.

Although causality cannot be definitively established, the data suggest that this rise was likely driven by a combination of factors. Programs such as Research4Life and publisher-specific 100% APC waivers played a central role in removing financial barriers for Ukrainian authors at a time when institutional support for publication costs was severely constrained.

In addition to formal waiver mechanisms, more flexible editorial policies, including expedited review processes and reduced documentation requirements, may have further contributed to increased accessibility. Finally, an overarching sense of international academic solidarity, coupled with an international drive to sustain Ukrainian scholarship under extraordinary conditions, likely helped Ukrainian authors overcome logistical and procedural obstacles to publish in high-visibility Open Access venues offered by the selected publishers.

*4.2. Alternative or complementary drivers*

While APC-waiver programs appear to have facilitated publication access for Ukrainian authors during wartime, they are unlikely to be the sole explanation for the observed increase in Gold OA output. Other mechanisms of support, including emergency research grants, international partnerships, and humanitarian academic programs, may also have played a critical role. In particular, some Ukrainian researchers benefited from externally funded projects that either directly covered APCs or enabled participation in collaborative work where costs were absorbed by foreign institutions.

Moreover, previous studies indicate that authors from low- and middle-income countries, including Ukraine, often choose to publish in Gold OA journals even in the absence of full APC waivers. This behaviour is motivated not only by the need for open dissemination but also by strategic considerations such as journal indexation status, impact factor visibility, and rapid time-to-publication. As shown by Damaševičius and Zailskaitė-Jakštė (2023), Ukrainian authors have increasingly favoured MDPI journals, which, despite not offering comprehensive APC discounts for Ukraine, continue to attract submissions due to their operational efficiency and international visibility. Nazarovets (2024) similarly documents a sustained increase in Gold OA publishing from Ukraine in high-output commercial journals, highlighting that waiver availability is a significant, but not exclusive, driver of publishing decisions. Notably, MDPI journals were not included in the present dataset, as they did not implement APC-waiver policies comparable to those of the selected five publishers. Their exclusion reflects the study's deliberate focus on analysing behavioural responses to the unique waiver interventions during the war.

These findings underscore the importance of viewing the recent growth in Ukrainian Gold OA publications through a multi-causal lens, acknowledging both structural incentives and individual-level publication strategies. Therefore, while this study demonstrates a marked increase in Ukrainian publishing within the five selected publishers' Gold OA journals, it does not aim to explain overall trends across the broader OA landscape, which may be shaped by additional or overlapping drivers.

### *4.3. Awareness and institutional support*

Although APC-waiver initiatives were introduced by multiple publishers in response to the wartime disruption of Ukrainian academia, their effectiveness depends not only on formal availability but also on institutional mediation and researcher awareness. Many Ukrainian institutions lack dedicated administrative units or trained personnel to assist researchers with navigating waiver application procedures, interpreting eligibility criteria, or advocating on behalf of authors. As a result, even where waivers exist, they may remain underutilized or inaccessible to the majority of potential beneficiaries. This issue is particularly relevant in the case of the five publishers analysed in this study, whose waiver policies may not have been widely communicated or operationalized at the institutional level in Ukraine. To improve the impact of such support mechanisms, Ukrainian institutions should consider introducing awareness-raising activities such as informational webinars, internal guidance pages, and dedicated contact points for researchers applying for APC waivers.

However, a deeper structural issue goes beyond the publication phase itself. Submitting an article to a Gold OA journal represents the final step in a long research cycle – one that requires funding, lab infrastructure, fieldwork capacity, and stable employment. For many Ukrainian scholars, especially those working in conflict-affected regions or under resource-depleted conditions, the core challenge is not merely covering APCs, but producing research that meets the methodological and empirical standards required by high-impact journals. In this context, APC discounts, while valuable, are insufficient in the absence of broader systemic reforms. These include sustained international support for post-war reconstruction, institutional integrity, academic freedom, and targeted investment in scientific infrastructure.

Thus, meaningful inclusion of Ukrainian researchers in the global Open Access ecosystem will require not only publisher-level waivers, but also comprehensive policies that address the full research lifecycle, from idea to publication, including improved awareness and institutional capacity to leverage existing publishing support mechanisms such as APC waivers.

### *4.4. Methodological limitations*

Several methodological limitations should be acknowledged when interpreting the findings of this study. First, it was not possible to directly verify whether individual publications were supported by APC-waivers. Publishers do not disclose waiver status at the article level, and such information is typically not captured in bibliometric databases. As a result, conclusions about the influence of waiver programs remain inferential, based on temporal correlations and aggregate patterns within the subset of journals included in the study. While some articles may include acknowledgements mentioning APC waivers, a systematic text-mining analysis of these sections was beyond the scope of this study, but may offer a promising avenue for future research.

Second, the WoS Core Collection does not reliably identify the corresponding author in its metadata exports. Although the dataset was filtered using the affiliation tag CU=Ukraine, this approach cannot fully confirm whether the first or corresponding author was Ukrainian in each case, especially in multi-authored international collaborations. Although the RP (reprint author) field is available for some records, its inconsistent presence across records and limited metadata coverage prevented systematic use in this analysis. Moreover, prior studies have shown that databases such as Dimensions and OpenAlex offer broader coverage of non-English-language journals and outputs from smaller or regional publishers (Visser et al., 2021; Basson et al., 2022; Culbert et al., 2024). While WoS was chosen for its reliable metadata structure and consistency in indexing, the exclusion of these alternative sources may have limited the scope of this study – particularly in disciplines underrepresented in WoS. Future research could employ multi-source triangulation to expand coverage and corroborate the observed trends.

Third, the analysis was restricted to fully Gold OA journals published by the five largest academic publishers. While this selection captures a substantial portion of the global APC-based Open Access ecosystem, it excludes diamond OA journals (Simard et al., 2024) and non-commercial platforms, which may play an important role in regions with constrained research funding. Future research could examine whether Ukrainian researchers increased submissions to such no-fee journals during the crisis.

Although Ukraine participates in several Transformative Agreements (TAs) via Electronic Information for Libraries (EIFL) not-for-profit organization[7], these do not involve the five major commercial publishers analysed in this study. As such, their potential influence on Gold OA trends is likely limited in this context. The decision to exclude hybrid journals, which are often covered by TAs, further supports the focus on Gold OA journals with clearly identifiable APC policies. It is worth noting that the emergency APC waivers introduced by these five publishers in response to the war represented the first time in history that a large share of Ukrainian researchers gained free access to publish in major international Gold OA journals. Prior to this, the Ukrainian state did not fund APCs, and institutional or grant-based support for publication costs was extremely limited.

Fourth, the study did not include imprints or subsidiary publishers unless they were explicitly labelled under one of the five selected publishers (Elsevier, SAGE, Springer Nature, Taylor & Francis, and Wiley) in the WoS metadata. As a result, journals published under separate branding, such as Hindawi (acquired by Wiley), were not counted in the parent publisher totals. This may lead to a partial underestimation of overall Gold OA output for each publisher, but it preserves consistency in the dataset by relying on the metadata as recorded in the database.

Finally, the study covers a six-year time frame (2019–2024), which allows for the identification of short-term trends, especially those related to crisis response. However, it remains to be seen whether the observed growth in Ukrainian Gold OA output is sustainable over the long term, or whether it reflects a temporary peak driven by exceptional circumstances.

Despite these limitations, the study offers a focused and empirically grounded assessment of Ukrainian researchers' engagement with a specific segment of the Gold OA ecosystem under conditions of crisis-driven publisher support.

## 5. Conclusions

This study demonstrates a marked increase in the number of publications by Ukrainian researchers in fully Gold OA journals published by the five largest academic publishers (Elsevier, SAGE, Springer Nature, Taylor & Francis, and Wiley) following the onset of the full-scale war in 2022. These publishers introduced 100% APC waivers specifically for Ukrainian authors, creating a unique opportunity to examine whether such policies influence publishing behaviour under crisis conditions.

[7] https://esac-initiative.org/about/transformative-agreements/agreement-registry/

While the growth in publication output within these journals is likely influenced by the availability of APC waivers, the evidence also points to a more complex constellation of drivers. These include international research grants, collaborative authorship with foreign institutions, and institutional pressures to maintain academic visibility through publication in indexed journals.

The findings highlight the potential of well-targeted publishing support mechanisms during emergencies. However, they also expose persistent structural inequalities embedded in the APC-based publishing model – inequalities that waivers alone cannot resolve. Greater transparency in waiver policies, proactive institutional mediation, and more effective dissemination of support opportunities are needed to maximize their impact.

Future research should build on these findings by expanding the analysis beyond commercial publishers, incorporating diamond and platinum OA models, and exploring the lived experiences of authors through qualitative methods. Such work will be essential for developing evidence-based policies that promote equity, resilience, and sustainability in global scholarly communication.

**Acknowledgements**

The author would like to thank all Ukrainian defenders for the possibility to finalize and publish this work. The author also gratefully acknowledges all publishers and scholars worldwide who have supported Ukrainian researchers in this time of need. Also, I am grateful to the editors and reviewers of the MetaROR platform for their insightful and helpful comments on the preprint version, which significantly contributed to improving this manuscript. https://doi.org/10.70744/MetaROR.179.2.ea

**Competing interests**

The author does not have any competing interests.

**Funding statement**

This study was not supported by any sponsor or funder.

**Data availability**

The dataset supporting the findings of this study is openly available via Zenodo at https://doi.org/10.5281/zenodo.15450551

**AI Contribution Disclosure**

The author used ChatGPT-5 for translation and editing.